\documentclass[prd,notitlepage,longbibliography,nofootinbib,superscriptaddress,onecolumn,preprintnumbers]{revtex4-2}
\usepackage[utf8]{inputenc}

\usepackage{bm}
\usepackage{comment} 
\usepackage[colorlinks=true,urlcolor=blue,anchorcolor=black,citecolor=blue,linkcolor=red,filecolor=black,menucolor=black,pagecolor=black,linktocpage=true,pdfproducer=medialab,pdfa=true]{hyperref}
\usepackage{graphicx}
\usepackage{amsmath,latexsym,amssymb,mathrsfs,ascmac,mathtools}
\usepackage{multirow}
\usepackage{braket}
\begin{document}

\title{Ultraviolet Behavior of the Wheeler–DeWitt Equation in Hořava–Lifshitz Gravity}

\author{Takamasa Kanai}
\email{kanai@kochi-ct.ac.jp}

\affiliation{Department of Social Design Engineering,
National Institute of Technology (KOSEN), Kochi College,
200-1 Monobe Otsu, Nankoku, Kochi, 783-8508, Japan}

\begin{abstract}
We investigate the quantum structure of black hole interiors in Hořava–Lifshitz gravity by analyzing the Wheeler–DeWitt equation in minisuperspace. Focusing on the ultraviolet regime, where higher-order spatial curvature terms dominate, we derive analytical solutions in this UV limit for both the original Hořava–Lifshitz action and its analytically continued counterpart. We study their behavior near the event horizon and the classical singularity, with particular attention to the interpretation of the wave function in terms of the annihilation-to-nothing scenario proposed in general relativity. In this paper, we have considered cases in which the two-dimensional spatial section is spherical, planar, or hyperbolic, as well as models with positive, negative, or vanishing cosmological constant. In all cases, we find that the terms dominating in the ultraviolet regime, together with the effects of the running scaling parameter, act to suppress the annihilation-to-nothing behavior. These results suggest that, at least within the range explored in this study, the characteristic annihilation-to-nothing behavior does not appear in the ultraviolet regime of Hořava–Lifshitz gravity, and provide a new perspective on the understanding of singularity resolution in quantum gravity.
\end{abstract}
\maketitle

\section{Introduction}
Understanding the quantum nature of spacetime inside black holes remains one of the central challenges in quantum gravity. In classical general relativity (GR), black hole interiors inevitably contain spacetime singularities where curvature invariants diverge and the classical description breaks down \cite{Penrose:1964wq,Hawking:1970zqf,Hawking:1973uf}. It is widely believed that such singularities should be resolved by quantum gravitational effects, but a complete and consistent description is still lacking.

One useful framework for probing quantum aspects of black hole interiors is the minisuperspace Wheeler–DeWitt (WDW) equation \cite{Halliwell:1989myn,Kiefer:2013jqa,Kiefer:2008sw,Kiefer:2025udf}. By restricting attention to highly symmetric configurations, the WDW equation allows one to study the quantum dynamics of geometry in a tractable manner. In particular, for spherically symmetric black holes, the interior region can be described by a Kantowski–Sachs–type minisuperspace \cite{Kantowski:1966te}, in which the WDW equation admits exact analytical solutions. Within this framework, Yeom and collaborators have proposed an intriguing interpretation of the WDW wave function: the so-called annihilation-to-nothing scenario, in which wave packets corresponding to classical geometries with opposite arrows of time annihilate inside the horizon, effectively resolving the singularity at the quantum level \cite{Bouhmadi-Lopez:2019kkt,Yeom:2019csm,Yeom:2021bpd,Brahma:2021xjy}.

While this picture has been shown to arise in GR-based analyses, it is important to ask whether such a mechanism is robust against ultraviolet (UV) modifications of gravity. General relativity is widely regarded as a low-energy effective theory, and near the classical singularity higher-derivative corrections are expected to play a crucial role \cite{Weinberg:1978kz,Donoghue:1993eb,Donoghue:1994dn,Burgess:2003jk}. Therefore, conclusions drawn purely within GR may not be reliable once genuine UV effects are taken into account  \cite{Kanai:2025mrx}.

Hořava–Lifshitz (HL) gravity provides a natural setting to address this issue \cite{Horava:2008ih,Horava:2009uw,Horava:2010zj}. By abandoning Lorentz invariance at high energies and introducing anisotropic scaling between space and time, HL gravity achieves power-counting renormalizability while recovering GR in the infrared (IR) limit. The theory is characterized by a dimensionless scaling parameter $\lambda$, which controls the relative weight of different kinetic terms and is expected to run under the renormalization group flow. As a result, HL gravity offers a well-motivated framework to study how UV modifications and running couplings affect the quantum dynamics of black hole interiors.

In this work, we investigate the Wheeler–DeWitt equation for black hole interiors within Hořava–Lifshitz gravity, focusing on the UV regime where higher-order spatial curvature terms dominate. We derive analytical solutions in the UV-dominated regime for both the original HL action and its analytically continued counterpart, and construct wave packets to study their physical interpretation. Particular attention is paid to the behavior of the wave function near the horizon and the classical singularity, as well as to the role played by the scaling parameter $\lambda$.

Our main result is that the annihilation-to-nothing scenario, which emerges in the general-relativistic limit $\lambda = 1$, is generically suppressed once the scaling parameter deviates from unity. In the ultraviolet-dominated regime of Hořava–Lifshitz gravity and the spherically symmetric case, irrespective of the presence or absence of a cosmological constant, the analytical solutions of the Wheeler–DeWitt equation exhibit markedly different asymptotic behaviors near the event horizon and near the classical singularity. As a consequence, wave packets propagating along different arrows of time may collide, but they do not undergo subsequent annihilation. This implies that the annihilation-to-nothing behavior is suppressed in the ultraviolet regime of Ho\v{r}ava–Lifshitz gravity.

This conclusion holds for both the original and the analytically continued actions. In the case of planar black holes, it is further supported by numerical analyses based on the analytical solutions of the Wheeler–DeWitt equation. Although our investigation is restricted to values of $\lambda$ close to the general-relativistic limit, our results suggest that the annihilation-to-nothing scenario is not a generic feature of ultraviolet-complete gravity theories, but rather a special property of the general-relativistic regime.

The paper is organized as follows. In Sec.~\ref{sec:review-wdw}, we review the minisuperspace WDW equation for black hole interiors and Yeom’s interpretation within GR. In Sec.~\ref{sec:HL-basics}, we introduce Hořava–Lifshitz gravity and derive the corresponding WDW equation in the UV regime. Analytical solutions and their physical interpretation are presented in Sec.~\ref{sphere case}. In Sec.~\ref{planar case}, we discuss extensions to planar black holes, and in Sec.~\ref{non constant} we consider models without a cosmological constant. Finally, Sec.~\ref{summary} summarizes our results and discusses their implications for singularity resolution in quantum gravity.

Throughout this paper, we work in natural units where $\hbar=c=16\pi G_N=1$.

\section{Minisuperspace Wheeler-DeWitt equation inside spherical black holes}
\label{sec:review-wdw}

In this section, we review the minisuperspace Wheeler-DeWitt (WDW) equation describing the interior of spherically symmetric black holes, focusing on aspects relevant for the subsequent analysis. A detailed derivation and a more extensive discussion can be found in Ref.~\cite{Kanai:2025mrx}, to which we refer the reader for technical details.

According to Birkhoff’s theorem \cite{Misner:1973prb}, any spherically symmetric vacuum solution of the Einstein equations is necessarily static and uniquely described by the $D$-dimensional Schwarzschild-Tangherlini metric in arbitrary dimensions \cite{Tangherlini:1963bw}. As a result, the interior region of a Schwarzschild black hole provides a representative and universal setting for studying minisuperspace quantum dynamics in spherical symmetry.

\subsection{Minisuperspace WDW equation}

Inside the event horizon, the spacetime geometry can be parametrized by a Kantowski-Sachs-type metric \cite{Kantowski:1966te}. Upon substituting this ansatz into the Einstein-Hilbert action and performing the minisuperspace reduction, the dynamics can be expressed in terms of two
canonical variables $(X,Y)$. After canonical quantization, the Hamiltonian constraint leads to the minisuperspace Wheeler-DeWitt equation
\begin{equation}
\label{eq:WDW-review}
\left(
\frac{\partial^2}{\partial X^2}
-
\frac{\partial^2}{\partial Y^2}
+
4 r_s^{2(D-3)} e^{2Y}
\right)
\Psi(X,Y)=0,
\end{equation}
where $r_s$ denotes a length-scale constant, identified with the horizon radius in the spherically symmetric classical solution, i.e., the Schwarzschild radius. Apart from an overall power of $r_s$, this equation is independent of the spacetime dimension.

In these coordinates, the asymptotic region $X,Y\to -\infty$ corresponds to the event horizon, while the limit $X\to \infty$ and $Y\to -\infty$ represents the classical spacetime singularity.

\subsection{Wave-packet solutions and Yeom’s interpretation}

Equation~\eqref{eq:WDW-review} admits a complete set of mode solutions of the form
\begin{equation}
\psi_k(X,Y)= e^{-ikX} K_{ik}\!\left(2 r_s e^{Y}\right),
\end{equation}
where $K_{ik}$ is the modified Bessel function of the second kind. General solutions can be constructed as superpositions over the continuous
momentum parameter $k$,
\begin{equation}
\Psi(X,Y)=\int_{-\infty}^{\infty} dk\, f(k)\,\psi_k(X,Y),
\end{equation}
with $f(k)$ specifying the wave packet.

For suitable choices of $f(k)$, the resulting probability density $|\Psi|^2$ exhibits a pronounced peak that closely follows the classical trajectory in minisuperspace. This behavior was interpreted by Yeom and collaborators \cite{Bouhmadi-Lopez:2019kkt,Yeom:2019csm,Yeom:2021bpd,Brahma:2021xjy} as an \emph{annihilation-to-nothing} process occurring inside the event horizon. In this picture, two classical spacetime branches with opposite arrows of time interfere destructively, leading to a vanishing probability density away from the classical trajectories.

Related analyses have shown that this qualitative interpretation persists for topological black holes described by scalar-type WDW equations, while a Dirac-type formulation yields a modified version of the annihilation picture \cite{Kan:2021yoh,Kan:2021fmw,Kan:2022ism}.

The purpose of the present review is not to reproduce these results in detail, but to set the stage for investigating how this picture is altered once higher-derivative and quantum corrections are taken into account. As general relativity should be regarded as a low-energy effective theory,
such corrections are expected to play a crucial role in the resolution of classical singularities. In the following sections, we therefore extend this framework by incorporating classical higher-derivative terms and genuinely quantum curvature-squared effects, and examine their impact on the interior wave function and its interpretation.

\section{Ho\v{r}ava--Lifshitz gravity: basic formulation}
\label{sec:HL-basics}
In this section, we briefly summarize the basic formulation of Ho\v{r}ava-Lifshitz (HL) gravity \cite{Horava:2008ih,Horava:2009uw,Horava:2010zj} that will be relevant for our analysis. For simplicity, we work within the minimal version of the theory and focus on its essential structural features. HL gravity can be further categorized into “projectable” and “nonprojectable” versions, depending on whether the lapse function is assumed to depend only on time or also on spatial coordinates \cite{Sotiriou:2009gy,Sotiriou:2009bx,Jacobson:2010iu,Blas:2009qj,Blas:2010hb}. In this work,  we restrict our attention to configurations that belong to the projectable version of HL gravity. Within this setting, we further impose the Kantowski--Sachs symmetry on the spacetime metric to describe the black hole interior.

HL gravity is formulated with a preferred foliation of spacetime by constant-time hypersurfaces. Accordingly, the spacetime metric is written in the Arnowitt-Deser-Misner (ADM) form \cite{Misner:1973prb,Kiefer:2025udf}
\begin{equation}
 ds^2 = -N^2 dt^2 + h_{ij}(dx^i + N^i dt)(dx^j + N^j dt),
\end{equation}
where $N(t,\mathbf{x})$ is the lapse function, $N^i(t,\mathbf{x})$ is the shift vector, and $h_{ij}(t,\mathbf{x})$ denotes the induced spatial metric.

The presence of a preferred foliation implies that the symmetry of the theory is reduced compared to general relativity. Specifically, HL gravity is invariant under \emph{foliation-preserving diffeomorphisms},
\begin{equation}
 t \to t'(t), \qquad
 x^i \to x'^i(t,\mathbf{x}),
\end{equation}
which consist of time reparametrizations and time-dependent spatial diffeomorphisms. This restricted symmetry allows the inclusion of higher-order spatial derivative terms without introducing higher-order time derivatives.

The extrinsic curvature associated with the spatial hypersurfaces is defined as
\begin{equation}
 K_{ij} = \frac{1}{2N}
 \left(
 \dot h_{ij} - D_i N_j - D_j N_i
 \right),
\end{equation}
where a dot denotes a derivative with respect to the preferred time coordinate $t$, and $D_i$ is the covariant derivative compatible with
$h_{ij}$. Its trace is denoted by $K \equiv h^{ij}K_{ij}$.

The action of HL gravity can then be written as
\begin{align}
\label{Horava action}
S = \int dt\, d^3x \sqrt{h}\Biggl[
&\frac{2}{\kappa^2}\left(K_{ij}K^{ij}-\lambda K^2\right)
-\frac{\kappa^2}{2w^4}C_{ij}C^{ij}
+\frac{\kappa^2\mu}{2w^2}\epsilon^{ijl}R_{il}D_j R^k_{\ l} \nonumber\\
&-\frac{\kappa^2\mu^2}{8}R_{ij}R^{ij}
+\frac{\kappa^2\mu^2}{8(1-3\lambda)}
\left(
\frac{1-4\lambda}{4}R^2
+\Lambda_W R
-3\Lambda_W^2
\right)
\Biggr],
\end{align}
where $\kappa$, $\lambda$, $w$, $\mu$, and $\Lambda_W$ are constant parameters of the theory. Here $R_{ij}$ and $R$ denote the Ricci tensor and Ricci scalar constructed from the spatial metric $h_{ij}$, respectively, and $C_{ij}$ is the Cotton tensor defined by
\begin{equation}
C^{ij}
\equiv
\epsilon^{ikl} D_k
\left(
R^j_{\ l} - \frac{1}{4}R\,\delta^j_{\ l}
\right).
\end{equation}

The inclusion of higher-order spatial curvature terms improves the ultraviolet behavior of the theory and renders it power-counting
renormalizable. This improvement is achieved at the cost of explicitly breaking local Lorentz invariance at high energies, while Lorentz symmetry is expected to be approximately recovered in the infrared limit $\lambda \to 1$.

In the ultraviolet (UV) limit, the dynamics of the theory is dominated by the highest-order spatial derivative terms in the action, in particular by the Cotton tensor term $C_{ij}C^{ij}$ and the quadratic curvature contributions. In this regime, Lorentz invariance is explicitly broken, and both classical and quantum solutions can differ drastically from those of general relativity. The coupling constant $\lambda$, which controls the relative weight of $K_{ij}K^{ij}$ and $K^{2}$ in the kinetic term, plays a crucial role in determining the ultraviolet structure of the theory and is expected to run under the renormalization group flow.

In contrast, in the infrared (IR) limit general relativity is recovered provided that the running coupling constant approaches its relativistic value $\lambda = 1$. In this limit, the effective speed of light, Newton’s constant, and cosmological constant emerge as
\begin{equation}
c = \frac{\kappa^{2}\mu}{4}\sqrt{\frac{\Lambda_{W}}{1-3\lambda}}, \qquad
G_{N} = \frac{\kappa^{2}}{32\pi c}, \qquad
\Lambda = \frac{3}{2}\Lambda_{W}.
\end{equation}

It is also important to note that, according to the above expressions, for $\lambda > 1/3$ the cosmological constant necessarily becomes negative, which is incompatible with observations. However, as pointed out in Ref.~\cite{Lu:2009em}, this problem can be resolved by performing an analytic continuation of the coupling constants, $\mu \to i\mu$ and $w^{2} \to -iw^{2}$. This procedure effectively changes the sign of the potential term in the action~(\ref{Horava action}), while leaving the kinetic term unchanged. As a consequence, the emergent speed of light becomes
\begin{equation}
c = \frac{\kappa^{2}\mu}{4}\sqrt{\frac{\Lambda_{W}}{3\lambda - 1}},
\end{equation}
and the resulting cosmological constant is positive for $\lambda > 1/3$, as desired. Throughout this paper, we work in units where the speed of light and Newton’s constant are set to unity, $c = 1$ and $16\pi G = 1$.

Due to the preferred foliation, the Hamiltonian formulation of HL gravity closely resembles that of general relativity. The canonical momenta conjugate to the spatial metric are
\begin{align}
\Pi^{ij} &= \frac{2\sqrt{h}}{\kappa^{2}} G^{ijkl} K_{kl},\\
G^{ijkl} &= \frac{1}{2}(g^{ik}g^{jl}+g^{il}g^{jk})-\lambda g^{ij}g^{kl},
\end{align}
where $G^{ijkl}$ is the generalized DeWitt metric. 

\subsection{Hamiltonian constraint}
As already pointed out in Ref.~\cite{Horava:2008ih}, a Hamiltonian formulation of gravity with anisotropic scaling is particularly natural due to the $3+1$ decomposition of spacetime inherent to the theory. In this framework, the spatial metric components become canonical variables, and the dynamics is conveniently described in terms of their conjugate momenta. As in general relativity, the Hamiltonian can be expressed as a linear combination of constraints associated with the lapse and shift functions, namely,
\begin{align}
H=\int dx^3(N\mathcal{H}_\perp+N^i\mathcal{H}_i),
\end{align}
where $\mathcal{H}_\perp$ and $\mathcal{H}_i$ are the Hamiltonian constraint and the momentum constraint, respectively. They are given by
\begin{align}
\mathcal{H}_\perp&=\frac{\kappa^2}{2\sqrt{h}}\Pi^{ij}\mathcal{G}_{ijkl}\Pi^{kl}+\frac{2\sqrt{h}}{\kappa^2}\mathcal{V},\\
\mathcal{H}_i&=-2D_j\Pi^{ij},
\end{align}
where $\mathcal{G}_{ijkl}$ denotes the inverse of the generalized DeWitt metric, and $\mathcal{V}$ represents the potential term, which are defined by
\begin{align}
\mathcal{G}_{ijkl}=&\frac{1}{2}(g^{ik}g^{jl}+g^{il}g^{jk})-\frac{\lambda}{3\lambda-1} g^{ij}g^{kl},\\
\label{potencial}
\mathcal{V}=&\frac{\kappa^4}{4w^4}C_{ij}C^{ij}-\frac{\kappa^4\mu}{4w^2}\epsilon^{ijl}R_{il}D_{j}R^j_{\ k}+\frac{\kappa^4\mu^2}{16}R_{ij}R^{ij}\nonumber\\
&-\frac{\kappa^4\mu^2}{16(1-3\lambda)}\left(\frac{1-4\lambda}{4}R^2+\Lambda_WR-3\Lambda_W^2\right).
\end{align}
It is worth noting that the Hamiltonian and momentum constraints in HL gravity preserve the same formal structure as those of general relativity in the ADM formulation. However, when the projectability condition is imposed, the Hamiltonian constraint becomes nonlocal, and the resulting constraint algebra differs slightly from that of general relativity~\cite{Horava:2008ih}. In this case, the relevant Hamiltonian constraint is given by its spatially integrated form,
\begin{equation}
\mathcal{H}_0 \equiv \int d^3x\, \mathcal{H}_{\perp}.
\end{equation}

Quantization can then be carried out following Dirac’s prescription by promoting the canonical variables to operators satisfying the usual commutation relations and imposing the constraints on the quantum state. Accordingly, the quantum version of the integrated Hamiltonian constraint $\mathcal{H}_{0}$, together with the momentum constraints $\mathcal{H}_{i}$, plays the role of the WDW equations,
\begin{equation}
\mathcal{H}_0\Psi = 0, \qquad \mathcal{H}_{i}\,\Psi = 0.
\end{equation}

In Sec.~\ref{sphere case}, we apply this quantization procedure to the Kantowski–Sachs cosmological model within the minimal version of HL gravity defined by the action~(\ref{Horava action}), thus exploring the quantum properties of the theory.

\section{Spherical and Hyperbolic Black Hole Interiors}
\label{sphere case}
Since general relativity, regarded as an effective field theory of gravity, is not expected to resolve spacetime singularities, we investigate how Yeom’s annihilation to nothing interpretation is realized within HL gravity, which is considered a candidate for a UV-complete theory of gravity.

We begin by considering the spherically symmetric black hole geometry, which provides the simplest and most widely studied setting for investigating the quantum structure of black hole interiors. Spherical symmetry allows for a significant simplification of the gravitational dynamics while retaining the essential features of horizon formation and classical singularities. In the context of HL gravity, spherically symmetric black hole solutions have been extensively analyzed and serve as a natural starting point for exploring ultraviolet modifications of black hole interiors within a minisuperspace framework \cite{Kehagias:2009is}. In what follows, we focus on this class of geometries and analyze the corresponding WDW equation. The metric in this case is given by
\begin{align}
\label{Kantowski metric}
ds^2 = - e^{-2X(t)+4Y(t)} dt^2 + e^{2X(t)} dr^2 + r_s^2 e^{-2X(t)+2Y(t)}d\Omega^2_2 ,
\end{align}
where  $d\Omega^2_2$ is the line element of the $2$ dimensional sphere with unit radius, and $r_s$ is a constant of dimension length, identified with the horizon radius in the spherically symmetric classical solution.

For this metric ansatz, the action reduces to
\begin{align}
S\propto&\int dt\Bigl(\frac{1}{2}(3-\lambda)\dot{X}^2-2(1-\lambda)\dot{X}\dot{Y}+(1-2\lambda)\dot{Y}^2\nonumber\\
&\mp\frac{1-3\lambda}{\Lambda_W}\Bigl\{ 2\Lambda_Wr_s^2 e^{2Y}+(2\lambda-1)e^{2X}-3\Lambda_W^2r_s^4e^{-2X+4Y}\Bigr\}\Bigr).
\end{align}

Therefore, we obtain the canonical Hamiltonian
\begin{align}
H=&\Bigl((2\lambda-1)\Pi_X^2-2(\lambda-1)\Pi_X\Pi_Y+\frac{1}{2}(\lambda-3)\Pi_Y^2\nonumber\\
&\pm\frac{1-3\lambda}{\Lambda_W}\Bigl\{ 2\Lambda_Wr_s^2 e^{2Y}+(2\lambda-1)e^{2X}-3\Lambda_W^2r_s^4e^{-2X+4Y}\Bigr\}\Bigr),
\end{align}
where $\Pi_X$ and $\Pi_Y$ are the canonical conjugate momenta to $X$ and $Y$, respectively, given by
\begin{align}
\label{momenta x}
\Pi_X&=(3-\lambda)\dot{X}-2(1-\lambda)\dot{Y},\\
\label{momenta y}
\Pi_Y&=-2(1-\lambda)\dot{X}+2(1-2\lambda)\dot{Y}.
\end{align}
We quantize the system by promoting the canonical variables to operators satisfying the canonical commutation relations
\begin{align}
[X,\Pi_X] = i, \qquad [Y,\Pi_Y] = i,
\end{align}
with all other commutators vanishing. In the coordinate representation, the momenta are realized as
\begin{align}
\Pi_X \rightarrow -i \frac{\partial}{\partial X}, \qquad
\Pi_Y \rightarrow -i \frac{\partial}{\partial Y}.
\end{align}
Substituting these into the Hamiltonian constraint, the Wheeler--DeWitt equation follows. The relative sign between the kinetic and potential terms arises from $(-i)^2 = -1$.

To derive the WDW equation, we promote the canonical conjugate momenta to quantum operators. Then, the resulting WDW equation are
\begin{align}
&\Bigl((2\lambda-1)\frac{\partial^2}{\partial X^2}-2(\lambda-1)\frac{\partial^2}{\partial X\partial Y}+\frac{1}{2}(\lambda-3)\frac{\partial^2}{\partial Y^2}\nonumber\\
&\mp\frac{1-3\lambda}{\Lambda_W}\Bigl\{ 2\Lambda_Wr_s^2 e^{2Y}+(2\lambda-1)e^{2X}-3\Lambda_W^2r_s^4e^{-2X+4Y}\Bigr\}\Bigr)\Psi(X,Y)=0,
\end{align}
where the minus and plus signs in the potential, corresponding to the original and the analytically continued actions, respectively \cite{Obregon:2012bt}. These quantum equations differ significantly from the usual Wheeler-DeWitt equation~(\ref{eq:WDW-review}). However, for this particular model with $\lambda = 1$, they reduce to Eq.~(\ref{eq:WDW-review}) up to the first two terms of the potential in Eq.~(\ref{potencial}). Therefore, in the infrared limit the model reproduces the same behavior as that obtained in general relativity.

On the other hand, the potential contains an additional term arising from the higher-order contributions in the action~(\ref{Horava action}), which governs the ultraviolet behavior of the model. Since it is difficult to obtain analytical solutions for the full set of equations, we focus on the UV limit, where this last term dominates the potential. In this regime, the Wheeler-DeWitt equations reduce to
\begin{align}
\Bigl((2\lambda-1)\frac{\partial^2}{\partial X^2}-2(\lambda-1)\frac{\partial^2}{\partial X\partial Y}+\frac{1}{2}(\lambda-3)\frac{\partial^2}{\partial Y^2}\mp\frac{1}{\Lambda_W}(1-3\lambda)(2\lambda-1)e^{2X}\Bigr)\Psi(X,Y)=0.
\end{align}

To solve the Wheeler--DeWitt equation in the UV regime, we employ the separable ansatz
\begin{align}
\Psi(X,Y) = e^{-ikY}\,\phi(X).
\end{align}
Substituting this into Eq.~(XX), we obtain an ordinary differential equation for $\phi(X)$:
\begin{align}
(2\lambda-1)\phi''(X)
-2(\lambda-1)(-ik)\phi'(X)
+\frac{1}{2}(\lambda-3)(-k^2)\phi(X)
\mp \frac{1}{\Lambda_W}(1-3\lambda)(2\lambda-1)e^{2X}\phi(X)=0.
\end{align}

Introducing a rescaled function
\begin{align}
\phi(X) = e^{-\frac{\lambda-1}{2\lambda-1}ikX}\,\chi(X),
\end{align}
the first-derivative term is eliminated, and the equation reduces to
\begin{align}
\chi''(X) + \left(\alpha k^2 \mp \beta e^{2X}\right)\chi(X)=0,
\end{align}
where $\alpha$ and $\beta$ are constants depending on $\lambda$.

Defining a new variable $z=\gamma e^{X}$, the equation takes the standard form of the Bessel equation,
\begin{align}
z^2 \frac{d^2\chi}{dz^2} + z \frac{d\chi}{dz} + \left(z^2 - \nu^2\right)\chi = 0,
\end{align}
whose solutions are given by the Bessel functions $J_{\nu}(z)$ or modified Bessel functions $K_{\nu}(z)$, depending on the sign of the potential term.

The fundamental solutions are
\begin{align}
\label{PsiSolution1}
\psi_k(X,Y)= \left\{
\begin{array}{ll}
e^{-ikY}e^{-\frac{\lambda-1}{2\lambda-1}ikX}K_{\frac{\sqrt{3\lambda-1}}{\sqrt{2}(2\lambda-1)}ik}(\gamma e^{X}) & ({\rm original\ action}),\\
e^{-ikY}e^{-\frac{\lambda-1}{2\lambda-1}ikX}J_{\frac{\sqrt{3\lambda-1}}{\sqrt{2}(2\lambda-1)}ik}(\gamma e^{X}) & ({\rm analytically\ continued\ action}),
\end{array}
\right.
\end{align}
where $k$ is the separation constant associated with the plane wave expansion along the $Y$ direction, $K_{\nu}$ denotes the modified Bessel function of the second kind, $J_{\nu}$ denotes the Bessel function of the first kind, and the constant $\gamma$ is defined as
\begin{align}
\gamma\equiv\sqrt{\left|\frac{1-3\lambda}{\Lambda_W}\right|}.
\end{align}
Although $\psi_k(X,Y)$ is obtained by solving the equation in the UV limit, it is useful to compare it with the corresponding solution in general
relativity. In the special case $\lambda = 1$, the wave function reduces to a familiar form.
\begin{align}
\psi_k(X,Y)=e^{ikY}K_{ik}(\gamma e^{X}).
\end{align}
However, in the present case, the interchange of the roles of the minisuperspace variables $(X,Y)$ leads to a qualitatively different situation. In the UV-dominated WDW equation, the asymptotic behavior of the solutions near the horizon and near the classical singularity differs significantly. As a result, wave packets propagating along opposite time directions in minisuperspace do collide; however, they do not undergo subsequent annihilation. Consequently, the annihilation-to-nothing scenario is not realized in the present model.

The WDW equation admits the solutions (\ref{PsiSolution1}), whose behavior can be analyzed in the asymptotic regions corresponding to the event horizon and the black hole singularity.

In the limit $X=Y\rightarrow -\infty$, corresponding to the event horizon, the argument of the Bessel functions, $\gamma e^{X}$, approaches zero. In this regime, both solutions reduce to oscillatory, plane-wave like behavior, reflecting the regularity of quantum states at the horizon.

In contrast, the black hole singularity is approached in the limit $X\rightarrow +\infty$ and $Y\rightarrow -\infty$, where the argument $\gamma e^{X}$ diverges. For the solution associated with the original action, the wave function is governed by the modified Bessel function $K_{\nu}$, which decays exponentially for large arguments. Consequently, the wave function is strongly suppressed near the singularity, indicating a quantum avoidance of the classical singularity.

In contrast, for the analytically continued action, the solution is expressed in terms of the Bessel function $J_{\nu}$, which exhibits an oscillatory behavior at large arguments. In this case, the wave function remains oscillatory as the singularity is approached, leading to a qualitatively different quantum behavior.

Therefore, while both solutions are regular at the event horizon, their asymptotic behavior near the black hole singularity crucially depends on
the choice of the action. The exponential suppression associated with the modified Bessel function $K_{\nu}$ provides a natural mechanism for
singularity resolution in the ultraviolet regime of HL gravity.

Even when the two-dimensional spatial section is hyperbolic, the analysis differs from the spherically symmetric case only by the sign of the curvature. Since the dominant contributions in the ultraviolet regime arise from curvature-squared terms, the resulting behavior of the singularity resolution is therefore the same as in the spherically symmetric case.

\section{Planar Black hole interiors}
\label{planar case}
In addition to spherically symmetric geometries, HL gravity also admits black hole solutions with planar horizons \cite{Cai:2009pe}. It is therefore natural to investigate the quantum structure of planar black hole interiors within this framework. In the following, we analyze the corresponding WDW equation in minisuperspace.

We consider the following metric ansatz for a planar black hole geometry:
\begin{align}
ds^2 = - e^{-2X(t)+4Y(t)} dt^2 + e^{2X(t)} dr^2 + r_s^2 e^{-2X(t)+2Y(t)}d\bar{\Omega}^2_2 ,
\end{align}
where $d\bar{\Omega}^2_2$ is the line element of the flat $2$ dimensional torus.

By performing the same calculation as in the previous section, the action becomes
\begin{align}
S\propto\int dt\Bigl(\frac{1}{2}(3-\lambda)\dot{X}^2-2(1-\lambda)\dot{X}\dot{Y}+(1-2\lambda)\dot{Y}^2\pm3(1-3\lambda)\Lambda_Wr_s^4e^{-2X+4Y}\Bigr).
\end{align}
An interesting feature of the planar black hole is that the curvature of constant-time hypersurfaces vanishes, so that terms constructed from the spatial curvature do not contribute. As a result, even in the ultraviolet regime, the theory does not receive significant modifications relative to general relativity.

Therefore, we obtain the canonical Hamiltonian
\begin{align}
H=&\Bigl((2\lambda-1)\Pi_X^2-2(\lambda-1)\Pi_X\Pi_Y+\frac{1}{2}(\lambda-3)\Pi_Y^2\mp3(1-3\lambda)\Lambda_Wr_s^4e^{-2X+4Y}\Bigr),
\end{align}
where $\Pi_X$ and $\Pi_Y$ are the canonical conjugate momenta to $X$ and $Y$,respectively, which are defined as (\ref{momenta x}), (\ref{momenta y}). The WDW equation is
\begin{align}
\Bigl((2\lambda-1)\frac{\partial^2}{\partial X^2}-2(\lambda-1)\frac{\partial^2}{\partial X\partial Y}+\frac{1}{2}(\lambda-3)\frac{\partial^2}{\partial Y^2}\pm3(1-3\lambda)\Lambda_Wr_s^4e^{-2X+4Y}\Bigr)\Psi(X,Y)=0.
\end{align}
To render the potential term dependent on a single variable and to facilitate the analysis, we perform a change of variables $(X,Y)\rightarrow(Z,W)$, under which the WDW equation becomes
\begin{align}
\Bigl(\frac{1}{2}(25\lambda-19)\frac{\partial^2}{\partial Z^2}-20(\lambda-1)\frac{\partial^2}{\partial Z\partial W}+(8\lambda-11)\frac{\partial^2}{\partial W^2}\pm3(1-3\lambda)\Lambda_Wr_s^4e^{2W}\Bigr)\Psi(Z,W)=0,
\end{align}
where $Z=2X-Y$ and $W=-X+2Y$. In this coordinate system, the region $Z,W\rightarrow-\infty$ represents the event horizon, whereas the limit $Z\rightarrow\infty$ and $W\rightarrow-\infty$ corresponds to the spacetime singularity. The fundamental solution of the equation is
\begin{align}
\label{solution1}
\psi_k(Z,W)= \left\{
\begin{array}{ll}
e^{-ikZ}e^{-\frac{10(\lambda-1)}{8\lambda-11}ikW}K_{\frac{\sqrt{9(3\lambda-1)}}{\sqrt{2}(11-8\lambda)}ik}(\nu e^{W}) & ({\rm original\ action}),\\
e^{-ikZ}e^{-\frac{10(\lambda-1)}{8\lambda-11}ikW}J_{\frac{\sqrt{9(3\lambda-1)}}{\sqrt{2}|11-8\lambda|}ik}(\nu e^{W}) & ({\rm analytically\ continued\ action}),
\end{array}
\right.
\end{align}
where $k$ is the separation constant associated with the plane wave expansion along the $Z$ direction, and the constant $\nu$ is defined as
\begin{align}
\nu\equiv\sqrt{\left|\frac{3(1-3\lambda)\Lambda_Wr_s^4}{8\lambda-11}\right|}.
\end{align}
The roles of the minisuperspace variables coincide with those in general relativity. For $\lambda = 1$, the solution further reduces to the corresponding GR result, allowing one to realize the annihilation-to-nothing interpretation. It is also possible that the annihilation-to-nothing scenario may occur for other values of $\lambda$, and we investigate this possibility in what follows.

We impose as a boundary condition that the wave function reduces to a Gaussian wave packet peaked on a classical trajectory at the horizon, and investigate its subsequent evolution in order to discuss the resolution of the singularity and the annihilation-to-nothing scenario.

In the asymptotic limit $W\rightarrow-\infty$, the modified Bessel function and the ordinary Bessel function simplify considerably.
It can be written as
\begin{align}
K_{ik}(\nu e^W)&\simeq\frac{1}{2}\Bigl(\left(\frac{\nu}{2}\right)^{ik}\Gamma(-ik)e^{ikW}+\left(\frac{\nu}{2}\right)^{-ik}\Gamma(ik)e^{-ikW}\Bigr),\\
J_{ik}(\nu e^W)&\simeq\left(\frac{\nu}{2}\right)^{ik}\frac{1}{\Gamma(1+ik)}e^{ikW}.
\end{align}
This forms show that the wave function constructed from the modified Bessel function exhibits two dominant peaks along $Z=W$ and $Z=-W$, while the solution based on the ordinary Bessel function is characterized by a single peak along $Z=W$. These peaks of the modified Bessel function are interpreted respectively as near-horizon and near-singularity behaviors and the peak of the Bessel function is interpreted respectively as near-horizon . The wavefunctions eq. (\ref{solution1}) are 
\begin{align}
\Psi(Z,W)\Big|_{Z=W\to-\infty}
\simeq
\left\{
\begin{array}{ll}
\displaystyle
\int_{-\infty}^{\infty} dk\, \frac{f(k)}{2}
\begin{aligned}[t]
\Bigl[
&\left(\frac{\nu}{2}\right)^{ik}\Gamma(-ik)\,e^{-ik(Z-W)} \\
&+\left(\frac{\nu}{2}\right)^{-ik}\frac{1}{\Gamma(ik)}\,e^{-ik(Z+W)}
\Bigr]
\end{aligned}
&
({\rm original\ action}),
\\[2ex]
\displaystyle
\int_{-\infty}^{\infty} dk\,
f(k)\left(\frac{\nu}{2}\right)^{ik}
\frac{1}{\Gamma(1+ik)}\,e^{-ik(Z-W)}
&
({\rm analytically\ continued\ action}).
\end{array}
\right.
\end{align}
In the vicinity of the horizon, the Hořava--Lifshitz scaling parameter is expected to flow to the infrared fixed point, $\lambda \to 1$, where the theory effectively reproduces general relativity. In this region, the Wheeler--DeWitt equation reduces to its GR form, and we impose a Gaussian wave-packet boundary condition,
\begin{align}
\Psi(Z,W) = \int dk\, f(k)\, \psi_k(Z,W),
\end{align}
with $f(k)$ chosen such that the wave packet is localized near the horizon.

Away from the horizon, however, the running of $\lambda$ modifies the effective equation. Since the Wheeler--DeWitt equation is local in minisuperspace, the form of the solutions changes accordingly, leading to a transition from the Gaussian packet near the horizon to the asymptotic behavior shown in Fig.~XX.
\begin{align}
f(k)= \left\{
\begin{array}{ll}
A\frac{2^{1+ik}e^{-\frac{\sigma^2k^2}{2}}}{\Gamma(-ik)\nu^{ik}} & ({\rm original\ action}),\\
A\frac{2^{ik}\Gamma(1+ik)e^{-\frac{\sigma^2k^2}{2}}}{\nu^{ik}} & ({\rm analytically\ continued\ action}),
\end{array}
\right.
\end{align}
where $A$ is the normalization constant and $\sigma$ is the standard deviation of the pulse at $Z=W$. Then, the wave function is
\begin{align}
\label{solution2}
\Psi(Z,W)
=
\left\{
\begin{array}{ll}
\displaystyle
\int_{-\infty}^{\infty} dk\, A\,
e^{-\frac{\sigma^2 k^2}{2}}
e^{-ikZ}
e^{-\frac{10(\lambda-1)}{8\lambda-11}ikW}
\,
\mathcal{K}_k(W),
&
({\rm original\ action}),
\\[2ex]
\displaystyle
\int_{-\infty}^{\infty} dk\, A\,
e^{-\frac{\sigma^2 k^2}{2}}
e^{-ikZ}
e^{-\frac{10(\lambda-1)}{8\lambda-11}ikW}
\,
\mathcal{J}_k(W),
&
({\rm analytically\ continued\ action}),
\end{array}
\right.
\end{align}
where $\mathcal{K}_k(W)$ and $\mathcal{J}_k(W)$ are defined by
\begin{align}
\mathcal{K}_k(W)
&=
\frac{2^{1+ik}}{\Gamma(-ik)\nu^{ik}}
K_{\frac{\sqrt{9(3\lambda-1)}}{\sqrt{2}(11-8\lambda)}ik}
\!\left(\nu e^{W}\right),
\\[1ex]
\mathcal{J}_k(W)
&=
\frac{2^{ik}\Gamma(1+ik)}{\nu^{ik}}
J_{\frac{\sqrt{9(3\lambda-1)}}{\sqrt{2}|11-8\lambda|}ik}
\!\left(\nu e^{W}\right).
\end{align}
In this paper, we numerically evaluate the modulus squared of the wave function for different values of $\lambda$ and present the results in Fig.~1. In some regions, the plots appear to be truncated; this behavior is caused by rapid oscillations of the wave function, which exceed the numerical resolution, and does not indicate any divergence of the function itself. The wave function remains convergent in these regions.
The plots are shown mainly for values of $\lambda$ around the general-relativistic limit $\lambda=1$. When $\lambda=1$ deviates significantly from this value, the convergence of the numerical integration deteriorates, making it difficult to obtain reliable plots. For this reason, our analysis is restricted to a neighborhood of $\lambda=1$.

For $\lambda=1$, the wave packet constructed from the modified Bessel function (original action) follows the classical trajectory, and the wave components propagating with opposite arrows of time annihilate at $Z=0$. This behavior corresponds to the annihilation-to-nothing scenario proposed in Ref.~\cite{Bouhmadi-Lopez:2019kkt,Yeom:2019csm,Yeom:2021bpd,Brahma:2021xjy}. A crucial observation is that this annihilation behavior disappears once the parameter $\lambda$ is varied away from unity. Within the range of parameters explored in the present analysis, this indicates that the annihilation-to-nothing scenario does not occur when the Hořava–Lifshitz scaling parameter $\lambda$ runs. We emphasize, however, that our investigation has been restricted to the vicinity of $\lambda=1$. It remains possible that $\lambda$ admits a fixed point at some other value, at which the annihilation-to-nothing scenario could be realized.

For the Bessel-function case corresponding to the analytically continued action, we find that when $\lambda = 1$, the wave packet propagates from the horizon along a classical trajectory and exhibits a disappearance at a certain location. This disappearance behavior is no longer observed once the value of $\lambda$ is varied away from $\lambda = 1$. As in the case of the modified Bessel function, it remains possible that $\lambda$ admits a fixed point outside the range explored in this work, at which an annihilation-to-nothing–type scenario could be realized.

For $\lambda = 1.25$, the wave function disappears in the region where $W$ is negative, whereas for $\lambda = 0.75$, the wave function becomes localized around $Z = 0$. These results may indicate the emergence of a new, annihilation-like behavior of wave packets in the ultraviolet regime.

Therefore, for both the original action and the analytically continued action, we find that allowing the scaling parameter to run away from the general relativistic limit suppresses the annihilation-to-nothing–type behavior.
\begin{figure}[tbp]
  \centering

  \begin{minipage}[t]{0.45\linewidth}
    \centering
    \includegraphics[width=\linewidth]{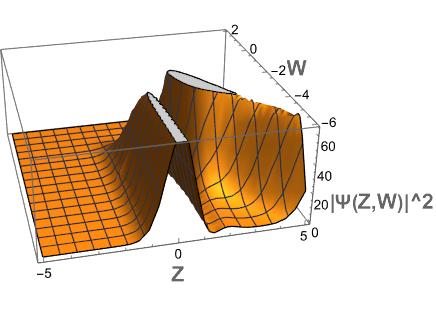}
    {\small (a) $\lambda=0.75$ (modified Bessel $K$)}
  \end{minipage}
  \hfill
  \begin{minipage}[t]{0.45\linewidth}
    \centering
    \includegraphics[width=\linewidth]{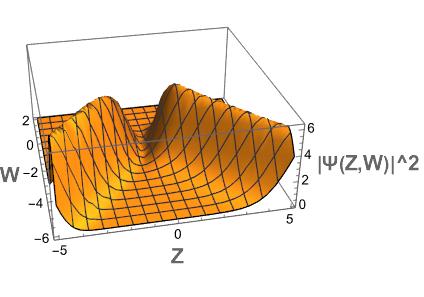}
    {\small (b) $\lambda=1$ (modified Bessel $K$)}
  \end{minipage}

  \vspace{5mm}

  \begin{minipage}[t]{0.45\linewidth}
    \centering
    \includegraphics[width=\linewidth]{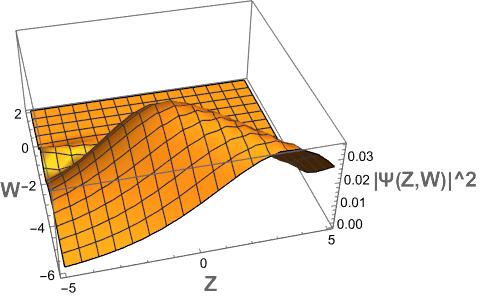}
    {\small (c) $\lambda=1.25$ (modified Bessel $K$)}
  \end{minipage}
  \hfill
  \begin{minipage}[t]{0.45\linewidth}
    \centering
    \includegraphics[width=\linewidth]{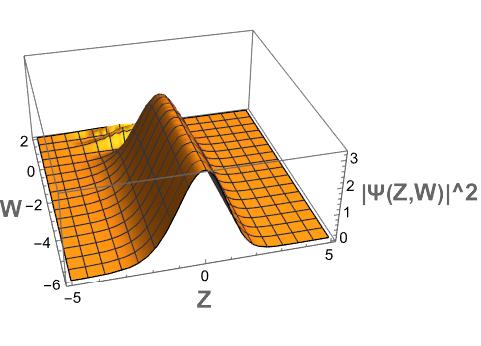}
    {\small (d) $\lambda=0.75$ (Bessel $J$)}
  \end{minipage}

  \vspace{5mm}

  \begin{minipage}[t]{0.45\linewidth}
    \centering
    \includegraphics[width=\linewidth]{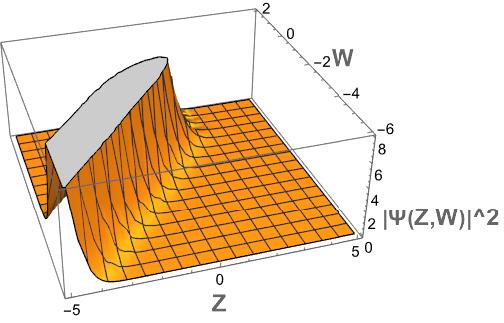}
    {\small (e) $\lambda=1$ (Bessel $J$)}
  \end{minipage}
  \hfill
  \begin{minipage}[t]{0.45\linewidth}
    \centering
    \includegraphics[width=\linewidth]{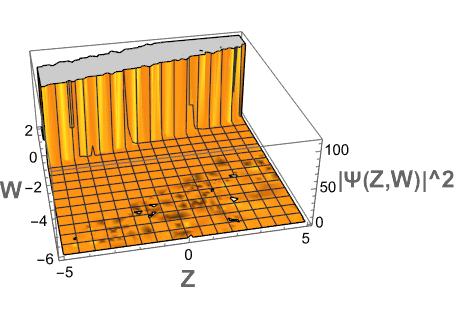}
    {\small (f) $\lambda=1.25$ (Bessel $J$)}
  \end{minipage}

  \caption{
  Modulus squared of the wave function \eqref{solution2} for different values of the HL parameter $\lambda$. The left (right) panels correspond to solutions expressed in terms of modified Bessel functions $K$ (Bessel functions $J$). In all panels, we set $A=\sigma=\nu/2=1$ and integrate the wave number over $k\in[-8,8]$.}
  \label{fig:wavefunction_lambda}
\end{figure}

\section{Non cosmological constant Black hole interiors}
\label{non constant}
In the previous section, we investigated the WDW equation in HL gravity in the presence of a nonvanishing cosmological constant. In this section, we turn to the spherically symmetric case without a cosmological constant. This extension allows us to clarify which features of the quantum dynamics are intrinsically tied to the presence of the cosmological constant and which persist as genuine ultraviolet effects of HL gravity. In particular, we examine how the absence of the cosmological constant modifies the structure of the minisuperspace WDW equation and the resulting behavior of the wave function near the horizon and the classical singularity.

We work within the $z=3$ HL gravity framework and truncate the potential at cubic order in the spatial curvature. No additional restrictions are imposed on the potential, and all coupling constants are treated as independent parameters. The HL gravity action is
\begin{align}
S = &\frac{2}{\kappa^2}\int dt\, d^3x\sqrt{-g}\Bigl\{K_{ij}K^{ij}-\lambda K^2+c_gR+L_{z>1}\Bigr],\\
L_{z>1}=&c_1D_iR_{jk}D^iR^{jk}+c_2D_iRD^iR+c_3R^j_{\ i}R^i_{\ k}R^k_{\ j}+c_4RR^j_{\ i}R^i_{\ j}+c_5R^3\nonumber\\
&+c_6R^j_{\ i}R^i_{\ j}+c_7R^2,
\end{align}
where $c_g,c_1,c_2,\cdots,c_7$ are the couplings of Ho\v{r}ava Lifshitz action treated as independent parameters. Although the bare cosmological constant is set to zero, quantum corrections are expected to generate an effective cosmological constant through renormalization flow (RG). For the black hole interior, spherical symmetry naturally leads to the Kantowski-Sachs metric. The HL gravity action evaluated for the Kantowski-Sachs metric~(\ref{Kantowski metric}) is
\begin{align}
S\propto&\int dt\Bigl(\frac{1}{2}(3-\lambda)\dot{X}^2-2(1-\lambda)\dot{X}\dot{Y}+(1-2\lambda)\dot{Y}^2+c_gr_s^2e^{2Y}\nonumber\\
&+\frac{1}{r_s^2}(c_3+2c_4+4c_5)e^{4X-2Y}+(c_6+2c_7)e^{2X}\Bigr).
\end{align}

We obtain the following Hamiltonian:
\begin{align}
H=&(2\lambda-1)\Pi_X^2-2(1-\lambda)\Pi_X\Pi_Y+\frac{1}{2}(\lambda-3)\Pi_Y^2-c_gr_s^2e^{2Y}\nonumber\\
&-\frac{1}{r_s^2}(c_3+2c_4+4c_5)e^{4X-2Y}-(c_6+2c_7)e^{2X}.
\end{align}
As discussed in the previous section, the ultraviolet behavior of the model is governed by the higher-order terms in the action. Accordingly, in the UV limit, the WDW equation reduces to
\begin{align}
&\Bigl\{(2\lambda-1)\frac{\partial^2}{\partial X^2}-2(1-\lambda)\frac{\partial^2}{\partial X\partial Y}+\frac{1}{2}(\lambda-3)\frac{\partial^2}{\partial Y^2}+\frac{1}{r_s^2}c_8e^{4X-2Y}\Bigr\}\Psi(X,Y)=0,
\end{align}
where $c_8=c_3+2c_4+4c_5$ is constant. Using the same change of variables $(X,Y)\rightarrow(Z,W)$ introduced previously to render the potential dependent on a single variable, the WDW equation becomes
\begin{align}
\Bigl(\frac{1}{2}(25\lambda-19)\frac{\partial^2}{\partial Z^2}-20(\lambda-1)\frac{\partial^2}{\partial Z\partial W}+(8\lambda-11)\frac{\partial^2}{\partial W^2}+\frac{1}{r_s^4}c_8e^{2Z}\Bigr)\Psi(Z,W)=0,
\end{align}
where the change of variables $(Z,W)$ defined by $Z=2X-Y$ and $W=-X+2Y$, the region $Z,W\rightarrow -\infty$ corresponds to the event horizon, while the
limit $Z\rightarrow\infty$ and $W\rightarrow -\infty$ represents the spacetime singularity. The fundamental solution of the equation is
\begin{align}
\psi_k(Z,W)= \left\{
\begin{array}{ll}
e^{-ikW}e^{-\frac{20(\lambda-1)}{25\lambda-19}ikZ}K_{\frac{\sqrt{-2(400\lambda^2-827\lambda+409)}}{(25\lambda-19)}ik}(\rho e^{Z}) & (c_8<0),\\
e^{-ikW}e^{-\frac{20(\lambda-1)}{25\lambda-19}ikZ}J_{\frac{\sqrt{-2(400\lambda^2-827\lambda+409)}}{(25\lambda-19)}ik}(\rho e^{Z}) & (c_8>0),
\end{array}
\right.
\end{align}
where the constant $\sigma$ is defined as
\begin{align}
\sigma\equiv\sqrt{\left|\frac{2c_8}{(25\lambda-19)r_s^4}\right|}.
\end{align}
This result yields a solution analogous to that obtained in the spherically symmetric case, exhibiting the same behavior with respect to singularity resolution. Even in this solution, wave packets propagating along opposite time directions in minisuperspace do collide; however, they do not undergo subsequent annihilation, and therefore the annihilation-to-nothing scenario is not realized in the present model.

Although the detailed form of the WDW equation differs in the absence of a cosmological constant, the asymptotic behavior of its solutions remains essentially unchanged. In particular, both solutions are regular at the event horizon, whereas their asymptotic behavior near the black hole singularity depends crucially on the choice of the action: the solution associated with $c_8>0$ is exponentially suppressed, while that corresponding to $c_8<0$ remains oscillatory.

These results demonstrate that the absence of the cosmological constant does not qualitatively alter the ultraviolet quantum behavior of the
black hole interior in Ho\v{r}ava-Lifshitz gravity.

\section{Summary and Discussion}
\label{summary}
In this paper, we have investigated the quantum structure of black hole interiors in Hořava–Lifshitz gravity by analyzing the Wheeler–DeWitt equation in minisuperspace. Focusing on the ultraviolet (UV) regime, where higher-order spatial curvature terms dominate the dynamics, we derived exact analytical solutions for both the original and the analytically continued actions and studied their physical implications.

Our analysis shows that, for both formulations, the resulting wave functions are regular at the event horizon. However, their asymptotic behavior near the classical singularity crucially depends on the choice of the action. In the case of the original action, the wave function is exponentially suppressed toward the singularity, indicating a quantum avoidance of the classical singularity. In contrast, the analytically continued action leads to oscillatory behavior, suggesting a qualitatively different quantum structure in the deep interior.

A central result of this work is that the interchange of minisuperspace variables inherent to the UV-dominated Wheeler–DeWitt equation leads to markedly different behaviors of the solutions near the horizon and near the classical singularity. As a consequence, wave packets propagating along different arrows of time do not annihilate, and the annihilation-to-nothing scenario—previously proposed as a mechanism for singularity resolution in general relativity—does not occur in the UV regime of Hořava–Lifshitz gravity. This result highlights a qualitative distinction between general relativity and its anisotropically scaled UV completion.

We have further examined the dependence of this behavior on the Hořava–Lifshitz scaling parameter $\lambda$. In the general-relativistic limit $\lambda=1$, the solutions reduce to the known GR results, and the annihilation-to-nothing interpretation can be recovered. However, once $\lambda$ is allowed to deviate from unity, corresponding to a running coupling under the renormalization group flow, the annihilation behavior is suppressed. This conclusion holds for both the original and the analytically continued actions and is supported by our numerical analysis of wave-packet evolution.

Although our numerical investigation has been restricted to values of $\lambda$ in the vicinity of the GR limit, our results suggest a clear tendency: running of the scaling parameter away from $\lambda=1$ acts to suppress annihilation-to-nothing–type behavior. It remains an intriguing open question whether the renormalization group flow of Hořava–Lifshitz gravity admits another fixed point at which the annihilation-to-nothing scenario could be realized. Exploring this possibility would require extending the analysis beyond the parameter range considered here.

We have also shown that these conclusions persist in planar black hole geometries and in models without an explicit cosmological constant. Despite differences in the detailed form of the Wheeler–DeWitt equation, the qualitative behavior of the wave functions—regularity at the horizon, action-dependent behavior near the singularity, and the absence of annihilation due to multiple effective time directions—remains unchanged. This robustness suggests that our results capture a generic feature of UV quantum dynamics in Hořava–Lifshitz gravity.

In summary, our work demonstrates that ultraviolet modifications of gravity encoded in Hořava–Lifshitz theory qualitatively alter the quantum dynamics of black hole interiors. While classical singularities can be avoided through exponential suppression of the wave function, the annihilation-to-nothing scenario appears to be a special feature of the general-relativistic limit rather than a generic prediction of UV-complete gravity theories. These findings provide new insight into the interplay between anisotropic scaling, quantum cosmology, and singularity resolution, and motivate further studies of renormalization group effects and more general minisuperspace models in Hořava–Lifshitz gravity.

\bibliography{references}
\bibliographystyle{JHEP.bst}

\end{document}